\documentclass[%
 reprint,
 amsmath,amssymb,
 aps,
nofootinbib
]{revtex4-2}

\usepackage{graphicx}%
\usepackage{dcolumn}%
\usepackage{bm}%

\usepackage{tikz}
\usepackage{physics}
\usetikzlibrary{arrows.meta}
\usepackage{mathtools}
\usepackage{sidecap}
\usepackage{xfrac}
\usepackage{slashed}
\usepackage{hyperref}
\usepackage{tikz-feynman}
\tikzset{
    ->-/.style={decoration={
  markings,
  mark=at position .6 with {\arrow{>}}},postaction={decorate}},
    -<-/.style={decoration={
  markings,
  mark=at position .5 with {\arrow{<}}},postaction={decorate}},
}
\usepackage{graphicx}
\usepackage[font=scriptsize]{subcaption}
\usepackage[font=small]{caption}
\usepackage{float}

\makeatletter
\newcommand{\lettercite}[2][R]{%
  \begingroup
    \renewcommand\@cite[2][]{[\textnormal{#1}]}%
    \cite{#2}%
  \endgroup
}
\makeatother

\begin{document}

\preprint{APS/123-QED}

\author{Zeno Capatti}
\author{Valentin Hirschi}%
\affiliation{%
Institute of Theoretical Physics \\
University of Bern \\
 Hochschulstrasse 4\\
 3012 Bern, Switzerland
}%

\date{June 27, 2025}%
\title{Comment to:\\``Rewording Theoretical Predictions at Colliders with Vacuum Amplitudes''}
\begin{abstract}
This comment highlights flaws in two central results presented by Ramírez-Uribe et al. 
in the titular letter \texttt{PRL 133.211901}~\cite{Ramirez-Uribe:2024rjg}.
Firstly, the authors incorrectly claim that integrands of differential scattering cross-sections are obtained as \emph{single} ``residues'' of vacuum amplitudes, although a sharp distinction of initial and final states requires to consider \emph{two} iterated discontinuities. 
Secondly, they wrongly state that non-pinched thresholds always locally match in such a differential representation.
\end{abstract}

\maketitle

\section{\label{sec:vacuum_diagrams}Residues of vacuum amplitudes}

\begin{figure}[H]
\centering
\begin{subfigure}[t]{0.155\textwidth}
\centering
\scalebox{0.65}{\begin{tikzpicture}

    \node[inner sep=0pt] (E1) {};
    \node[inner sep=0pt] (E2) [below = 2cm of E1] {};

    \node[inner sep=0pt] (A1) [right = 1cm of E1] {};
    \node[inner sep=0pt] (A2) [right = 1cm of E2] {};
    \node[inner sep=0pt] (U1) {};

    \node[inner sep=0pt] (C) [below right = 1cm and 2cm of U1] {};

    \node[inner sep=0pt] (B) [right = 1cm of C] {};

    \node[inner sep=0pt] (EE1) [right = 4cm of E1]{};
    \node[inner sep=0pt] (EE2) [right = 4cm of E2] {};

    \node[inner sep=0pt] (C1) [above right = 0.3cm and 0.5cm of A1] {};
    \node[inner sep=0pt] (C2) [below right = 0.3cm and 0.5cm of A2] {};

        \begin{feynman}
        \draw[fermion, thick, black, line width=0.4mm, momentum=\(p_1\)] (E1) to (A1);
     \draw[fermion, thick, black, line width=0.4mm] (A1) to (A2);
     \draw[fermion, thick, black, line width=0.4mm, reversed momentum=\(p_2\)] (A2) to (E2);

     \draw[gluon, thick, black, line width=0.4mm, momentum=\(k_1\)] (A1) to (C);
     \draw[gluon, thick, black, line width=0.4mm, reversed momentum=\(k_2\)] (C) to (A2);
     \draw[gluon, thick, black, line width=0.4mm] (C) to (B);

     \draw[fermion, thick, black, line width=0.4mm] (B) to (EE1);

     \draw[fermion, thick, black, line width=0.4mm] (EE2) to (B);

     \draw[-, thick, red, line width=0.4mm, out=-120, in=80] (C1) to (C2);
    	
    	\end{feynman}

 \path[draw=black, fill=black] (A1) circle[radius=0.075];
 \path[draw=black, fill=black] (A2) circle[radius=0.075];
 \path[draw=black, fill=black] (C) circle[radius=0.075];
 \path[draw=black, fill=black] (B) circle[radius=0.075];

\end{tikzpicture}}
\subcaption{
$\substack{\displaystyle p_{1,2}^0=|\mathbf{p}_{1,2}| \\ \displaystyle k_{1,2}^0=|\mathbf{k}_{1,2}|}$
}
\label{fig:interference}
\end{subfigure}
\begin{subfigure}[t]{0.155\textwidth}
\centering
\scalebox{0.8}{\begin{tikzpicture}

    \node[inner sep=0pt] (U1) {};
    \node[inner sep=0pt] (U2) [below = 2cm of U1] {};

    \node[inner sep=0pt] (UL) [below left = 1cm and 1cm of U1] {};
    \node[inner sep=0pt] (UR) [below right = 1cm and 1cm of U1] {};

    \node[inner sep=0pt] (C) [below = 1cm of U1] {};

    \node[inner sep=0pt] (C1) [above right = 0.3 and 0.5cm of U1] {};
    \node[inner sep=0pt] (C2) [below left = 2cm and 1cm of U1] {};

        \begin{feynman}
        \draw[fermion, thick, black, line width=0.4mm, quarter left, reversed momentum=\(p_2\)] (U1) to (UR);
     \draw[fermion, thick, black, line width=0.4mm, quarter left] (UR) to (U2);
     \draw[-, thick, black, line width=0.4mm, quarter left, momentum=\(p_1\)] (U2) to (UL);
     \draw[fermion, thick, black, line width=0.4mm, quarter left] (UL) to (U1);

     \draw[gluon, thick, black, line width=0.4mm, reversed momentum=\(k_2\)] (C) to (U1);
     \draw[gluon, thick, black, line width=0.4mm, reversed momentum=\(k_1\)] (C) to (UL);
     \draw[gluon, thick, black, line width=0.4mm] (C) to (UR);

     \draw[-, thick, red, line width=0.4mm, out=-120, in=80] (C1) to (C2);
    	
    	\end{feynman}

 \path[draw=black, fill=black] (UL) circle[radius=0.075];
 \path[draw=black, fill=black] (UR) circle[radius=0.075];
 \path[draw=black, fill=black] (C) circle[radius=0.075];
 \path[draw=black, fill=black] (U1) circle[radius=0.075];
 \path[draw=black, fill=black] (U2) circle[radius=0.025];

\end{tikzpicture}}
\subcaption{
$\substack{\displaystyle p_{1,2}^0=|\mathbf{p}_{1,2}| \\ \displaystyle k_{1,2}^0=-|\mathbf{k}_{1,2}|}$
}
\label{fig:vacuum_diag}
\end{subfigure}
\begin{subfigure}[t]{0.155\textwidth}
\centering
\scalebox{0.82}{\begin{tikzpicture}

    \node[inner sep=0pt] (U1) {};
    \node[inner sep=0pt] (U2) [below = 2cm of U1] {};

    \node[inner sep=0pt] (UL) [below left = 1cm and 1cm of U1] {};
    \node[inner sep=0pt] (UR) [below right = 1cm and 1cm of U1] {};

    \node[inner sep=0pt] (C) [below = 1cm of U1] {};

    \node[inner sep=0pt] (C1) [above right = 0.3 and 0.5cm of U1] {};
    \node[inner sep=0pt] (C2) [below left = 2cm and 1cm of U1] {};

    \node[inner sep=0pt] (L1) [above right = 0.7cm and -0.2cm of UL] {$e_6$};
    \node[inner sep=0pt] (L2) [above left = 0.45cm and 0.3cm of UR] {$e_2$};
    \node[inner sep=0pt] (L3) [above left = 0.4cm and 0.1cm of C] {$e_4$};
    \node[inner sep=0pt] (L4) [above left = 0.15cm and 0.4cm of C] {$e_3$};
    \node[inner sep=0pt] (L5) [below right = 0.2cm and 0.4cm of C] {$e_5$};
    \node[inner sep=0pt] (L5) [below = 1.1cm and 0.1cm of C] {$e_1$};

        \begin{feynman}
        \draw[fermion, thick, black, line width=0.4mm, quarter left, reversed momentum=\(\ell_2\)] (U1) to (UR);
     \draw[fermion, thick, black, line width=0.4mm, quarter left] (UR) to (U2);
     \draw[-, thick, black, line width=0.4mm, quarter left, momentum=\(\ell_1\)] (U2) to (UL);
     \draw[fermion, thick, black, line width=0.4mm, quarter left] (UL) to (U1);

     \draw[gluon, thick, black, line width=0.4mm] (U1) to (C);
     \draw[gluon, thick, black, line width=0.4mm, reversed momentum=\(\ell_3\)] (C) to (UL);
     \draw[gluon, thick, black, line width=0.4mm] (C) to (UR);

    	\end{feynman}

 \path[draw=black, fill=black] (UL) circle[radius=0.075];
 \path[draw=black, fill=black] (UR) circle[radius=0.075];
 \path[draw=black, fill=black] (C) circle[radius=0.075];
 \path[draw=black, fill=black] (U1) circle[radius=0.075];
 \path[draw=black, fill=black] (U2) circle[radius=0.025];

\end{tikzpicture}}
\subcaption{
Labelled vacuum graph.}
\label{fig:vacuum_diag_labelled}
\end{subfigure}
\caption{According to the authors of ref.~\cite{Ramirez-Uribe:2024rjg}, (a) is the residue of the cut in (b), but their kinematics are different.}
\end{figure}
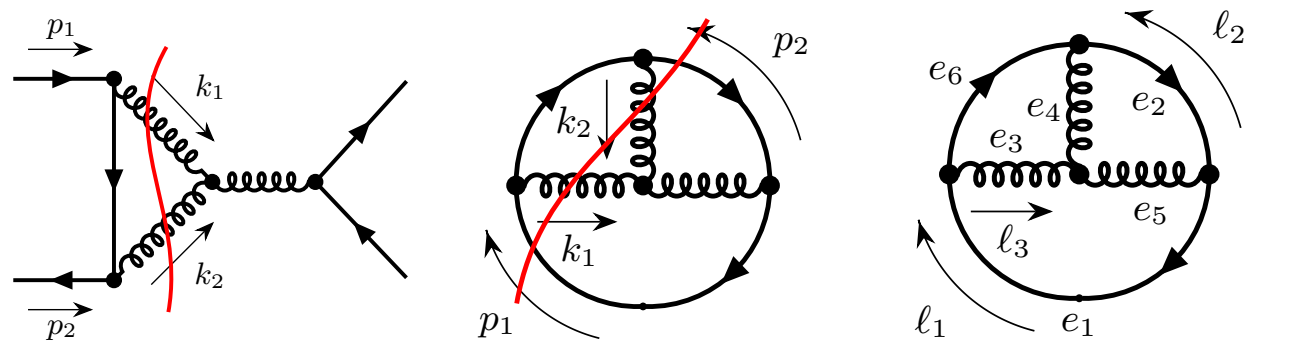

The authors of ref.~\cite{Ramirez-Uribe:2024rjg}\footnotemark[1] propose a development of the idea of deriving integrands for interference diagrams from vacuum graphs~\cite{Kinoshita:1962ur,Sterman:1993hfp,Capatti:2020xjc,Blazek:2021zoj,Capatti:2023omc}. At its core we find eqs.~(R.4) and (R.5), which can be shown to be incorrect by means of a tree-level counterexample: $q\bar{q}\rightarrow gg$.
We apply eq.~(R.5) to the computation of the residue of the cut of fig.~\ref{fig:vacuum_diag}. In the rest frame of $p_1+p_2$, we find\footnotemark[2]:
\begin{equation}
|\mathbf{k}_1||\mathbf{k}_2|\mathcal{A}_D^{(\Lambda,R)}\propto \frac{\tilde{t}^2 + \tilde{t} \tilde{u} + \tilde{u} (\tilde{s} + \tilde{u})}{\tilde{s}\tilde{t}},
\label{WrongResidue}
\end{equation}
where $\tilde{s}=2(|\mathbf{k}_1||\mathbf{k}_2|-\mathbf{k}_1\cdot \mathbf{k}_2)=s$,
$\tilde{t}=2(|\mathbf{p}_2||\mathbf{k}_2|+\mathbf{p}_2\cdot \mathbf{k}_2)$ and
$\tilde{u}=2(|\mathbf{p}_2||\mathbf{k}_1|+\mathbf{p}_2\cdot \mathbf{k}_1)$.
Note that $\tilde{s}+\tilde{t}+\tilde{u}\neq 0$. The correct result for the interference diagram of fig.~\ref{fig:interference} corresponds to eq.~\eqref{WrongResidue} but substituting the tilded Mandelstam invariants with
 their original expressions: $t=2(-|\mathbf{p}_2||\mathbf{k}_{2}|+\mathbf{p}_2\cdot \mathbf{k}_{2})$ and $u=2(-|\mathbf{p}_2||\mathbf{k}_{1}|+\mathbf{p}_2\cdot \mathbf{k}_{1})$.
Their mistake originates from the energy flow associated with the cut of the vacuum diagram, which is incompatible with that of the interference diagram. Ramírez-Uribe et al. appear to be partially aware of this fact, which explains their adjustment of the energy sign in eq.~(R.8) ($ab\rightarrow \bar{a}\bar{b}$) in order to obtain the correct phase-space integration measure. Even so, eq.~(R.5) yields integrands with incorrect numerators and denominators for any 
diagram with singularities that are \emph{not} s-channel, making it unsuitable for any genuine $2\rightarrow N$ process. 
Fixing this mistake, which affects energies and spatial momenta differently, requires significant conceptual novelties~\cite{Capatti:2020xjc,Capatti:2023omc,Capatti:2025gqj}.

\newpage
\section{\label{sec:differential}Threshold matching}
The authors claim that ``\emph{threshold singularities match locally among the different phase-space residues}'' and that ``\emph{the local matching of thresholds
requires phase-space residues with the same number of external particles and is not restricted by kinematical cuts.}''. These statements are incorrect. Among other things, they would imply the observable-independent cancellation of Glauber modes, and make obsolete  a wealth of literature concerning threshold regularisation~\cite{Soper:1999xk,Gong:2008ww,Becker:2012bi,Buchta:2015wna,Capatti:2019edf,Kermanschah:2021wbk,Kermanschah:2022mpb,Vicini:2024ecf}.

\begin{figure}[H]
\centering
\begin{subfigure}[h]{0.1\textwidth}
\centering
\vspace{0.000\textheight}
\scalebox{0.60}{\begin{tikzpicture}

    \node[inner sep=0pt] (U1) {};
    \node[inner sep=0pt] (U2) [below = 2cm of U1] {};

    \node[inner sep=0pt] (UL) [below left = 1cm and 1cm of U1] {};
    \node[inner sep=0pt] (UR) [below right = 1cm and 1cm of U1] {};

    \node[inner sep=0pt] (C) [below = 1cm of U1] {};

    \node[inner sep=0pt] (CU) [above = 0.4cm of C] {};

    \node[inner sep=0pt] (CD) [below = 0.4cm of C] {};

    \node[inner sep=0pt] (C1) [above right = 0. and 1cm of U1] {};
    \node[inner sep=0pt] (C2) [below left = 2cm and 1cm of U1] {};

        \begin{feynman}
        \draw[dashed, thick, black, line width=0.4mm, quarter left] (U1) to (UR);
     \draw[fermion, thick, black, line width=0.4mm, quarter left] (UR) to (U2);
     \draw[gluon, thick, black, line width=0.4mm, quarter right
     ] (UL) to (U2);
     \draw[fermion, thick, black, line width=0.4mm, quarter left] (UL) to (U1);

     \draw[fermion, thick, black, line width=0.4mm] (U1) to (CU);
     \draw[fermion, thick, black, line width=0.4mm] (CU) to (UL);
     \draw[fermion, thick, black, line width=0.4mm] (CD) to (UR);

     \draw[fermion, thick, black, line width=0.4mm] (U2) to (CD);

     \draw[gluon, thick, black, line width=0.4mm] (CD) to (CU);

     \draw[-, thick, red, line width=0.4mm, out=-110, in=70] (C1) to (C2);
    	
    	\end{feynman}

 \path[draw=black, fill=black] (UL) circle[radius=0.075];
 \path[draw=black, fill=black] (UR) circle[radius=0.075];
 \path[draw=black, fill=black] (CU) circle[radius=0.075];
 \path[draw=black, fill=black] (CD) circle[radius=0.075];
 \path[draw=black, fill=black] (U1) circle[radius=0.075];
 \path[draw=black, fill=black] (U2) circle[radius=0.025];

\end{tikzpicture}}
\vspace{0.000\textheight}
\subcaption{}
\label{fig:vacuumHtogg}
\end{subfigure}
\hspace{-0.01\textwidth}\begin{subfigure}[h]{0.13\textwidth}
\centering
\vspace{0.007\textheight}
\scalebox{0.66}{\begin{tikzpicture}

    \node[inner sep=0pt] (E1) {};
    \node[inner sep=0pt] (A1) [right = 0.5cm of E1] {};
    \node[inner sep=0pt] (A2) [above right = 0.5 and 0.7cm of A1] {};
    \node[inner sep=0pt] (A3) [below right = 0.5 and 0.7cm of A1] {};
    \node[inner sep=0pt] (B2) [right = 0.7 of A2] {};
    \node[inner sep=0pt] (B3) [right = 0.7 of A3] {};
    \node[inner sep=0pt] (B1) [right = 2.1cm of A1] {};

    \node[inner sep=0pt] (E2) [right = 0.5cm of B1] {};

     \node[inner sep=0pt] (C1L) [above right = 0.7cm and 0.45cm of A1] {};
     \node[inner sep=0pt] (C2L) [below right = 0.7cm and 0.45cm of A1] {};

     \node[inner sep=0pt] (C1) [above right = 0.7cm and 1.05cm of A1] {};
     \node[inner sep=0pt] (C2) [below right = 0.7cm and 1.05cm of A1] {};

     \node[inner sep=0pt] (C1R) [above right = 0.7cm and 1.65cm of A1] {};
     \node[inner sep=0pt] (C2R) [below right = 0.7cm and 1.65cm of A1] {};

        \begin{feynman}
     \draw[dashed, thick, black, line width=0.35mm] (E1) to (A1);
     \draw[dashed, thick, black, line width=0.35mm] (E2) to (B1);
     \draw[fermion, thick, black, line width=0.35mm] (A2) to (A1);
     \draw[fermion, thick, black, line width=0.35mm] (A1) to (A3);
     \draw[fermion, thick, black, line width=0.35mm] (A3) to (A2);
     \draw[fermion, thick, black, line width=0.35mm] (B1) to (B2);
     \draw[fermion, thick, black, line width=0.35mm] (B2) to (B3);
     \draw[fermion, thick, black, line width=0.35mm] (B3) to (B1);
     \draw[gluon, thick, black, line width=0.35mm] (A2) to (B2);
     \draw[gluon, thick, black, line width=0.35mm] (A3) to (B3);

     \draw[dashed, thick, blue, line width=0.35mm] (C1L) to (C2L);
     \draw[dashed, thick, blue, line width=0.35mm] (C1R) to (C2R);
     \draw[-, thick, red, line width=0.35mm] (C1) to (C2);
    	
    	\end{feynman}

 \path[draw=black, fill=black] (A1) circle[radius=0.075];
 \path[draw=black, fill=black] (A2) circle[radius=0.075];
 \path[draw=black, fill=black] (A3) circle[radius=0.075];
 \path[draw=black, fill=black] (B1) circle[radius=0.075];
 \path[draw=black, fill=black] (B2) circle[radius=0.075];
 \path[draw=black, fill=black] (B3) circle[radius=0.075];

\end{tikzpicture}}
\vspace{-0.0085\textheight}
\subcaption{}
\label{fig:interferenceHtogg}
\end{subfigure}
\hspace{0.0\textwidth}\begin{subfigure}[h]{0.24\textwidth}
\centering
\vspace{0.0085\textheight}
\scalebox{0.61}{\begin{tikzpicture}

    \node[inner sep=0pt] (E1) {};
    \node[inner sep=0pt] (A1) [right = 0.5cm of E1] {};
    \node[inner sep=0pt] (A2) [above right = 0.5 and 0.7cm of A1] {};
    \node[inner sep=0pt] (A3) [below right = 0.5 and 0.7cm of A1] {};
    \node[inner sep=0pt] (B2) [right = 0.7 of A2] {};
    \node[inner sep=0pt] (B3) [right = 0.7 of A3] {};
    \node[inner sep=0pt] (B1) [right = 2.1cm of A1] {};

    \node[inner sep=0pt] (E2) [right = 0.5cm of B1] {};

     \node[inner sep=0pt] (C1L) [above right = 0.7cm and 0.45cm of A1] {};
     \node[inner sep=0pt] (C2L) [below right = 0.7cm and 0.45cm of A1] {};

     \node[inner sep=0pt] (C1) [above right = 0.7cm and 1.05cm of A1] {};
     \node[inner sep=0pt] (C2) [below right = 0.7cm and 1.05cm of A1] {};

     \node[inner sep=0pt] (C1R) [above right = 0.7cm and 1.85cm of A1] {};
     \node[inner sep=0pt] (C2R) [below right = 0.7cm and 1.65cm of A1] {};

        \begin{feynman}
     \draw[dashed, thick, black, line width=0.35mm] (E1) to (A1);
     \draw[dashed, thick, black, line width=0.35mm] (E2) to (B1);
     \draw[fermion, thick, black, line width=0.35mm] (A2) to (A1);
     \draw[fermion, thick, black, line width=0.35mm] (A1) to (A3);
     \draw[fermion, thick, black, line width=0.35mm] (A3) to (A2);
     \draw[fermion, thick, black, line width=0.35mm] (B1) to (B2);
     \draw[fermion, thick, black, line width=0.35mm] (B2) to (B3);
     \draw[fermion, thick, black, line width=0.35mm] (B3) to (B1);
     \draw[gluon, thick, black, line width=0.35mm] (A2) to (B2);
     \draw[gluon, thick, black, line width=0.35mm] (A3) to (B3);

     \draw[-, thick, red, line width=0.35mm] (C1R) to (C2);
     \draw[-, dashed, blue, line width=0.45mm] (C1L) to (C2L);

    	\end{feynman}

 \path[draw=black, fill=black] (A1) circle[radius=0.075];
 \path[draw=black, fill=black] (A2) circle[radius=0.075];
 \path[draw=black, fill=black] (A3) circle[radius=0.075];
 \path[draw=black, fill=black] (B1) circle[radius=0.075];
 \path[draw=black, fill=black] (B2) circle[radius=0.075];
 \path[draw=black, fill=black] (B3) circle[radius=0.075];

\end{tikzpicture}} \scalebox{0.61}{\begin{tikzpicture}

    \node[inner sep=0pt] (E1) {};
    \node[inner sep=0pt] (A1) [right = 0.5cm of E1] {};
    \node[inner sep=0pt] (A2) [above right = 0.5 and 0.7cm of A1] {};
    \node[inner sep=0pt] (A3) [below right = 0.5 and 0.7cm of A1] {};
    \node[inner sep=0pt] (B2) [right = 0.7 of A2] {};
    \node[inner sep=0pt] (B3) [right = 0.7 of A3] {};
    \node[inner sep=0pt] (B1) [right = 2.1cm of A1] {};

    \node[inner sep=0pt] (E2) [right = 0.5cm of B1] {};

     \node[inner sep=0pt] (C1L) [above right = 0.7cm and 0.45cm of A1] {};
     \node[inner sep=0pt] (C2L) [below right = 0.7cm and 0.45cm of A1] {};

     \node[inner sep=0pt] (C1) [above right = 0.7cm and 1.05cm of A1] {};
     \node[inner sep=0pt] (C2) [below right = 0.7cm and 1.05cm of A1] {};

     \node[inner sep=0pt] (C1R) [above right = 0.7cm and 1.85cm of A1] {};
     \node[inner sep=0pt] (C2R) [below right = 0.7cm and 1.65cm of A1] {};

        \begin{feynman}
     \draw[dashed, thick, black, line width=0.35mm] (E1) to (A1);
     \draw[dashed, thick, black, line width=0.35mm] (E2) to (B1);
     \draw[fermion, thick, black, line width=0.35mm] (A2) to (A1);
     \draw[fermion, thick, black, line width=0.35mm] (A1) to (A3);
     \draw[fermion, thick, black, line width=0.35mm] (A3) to (A2);
     \draw[fermion, thick, black, line width=0.35mm] (B1) to (B2);
     \draw[fermion, thick, black, line width=0.35mm] (B2) to (B3);
     \draw[fermion, thick, black, line width=0.35mm] (B3) to (B1);
     \draw[gluon, thick, black, line width=0.35mm] (A2) to (B2);
     \draw[gluon, thick, black, line width=0.35mm] (A3) to (B3);

     \draw[-, thick, red, line width=0.35mm] (C1L) to (C2L);
      \draw[-, dashed, blue, line width=0.45mm] (C2) to (C1R);

    	\end{feynman}

 \path[draw=black, fill=black] (A1) circle[radius=0.075];
 \path[draw=black, fill=black] (A2) circle[radius=0.075];
 \path[draw=black, fill=black] (A3) circle[radius=0.075];
 \path[draw=black, fill=black] (B1) circle[radius=0.075];
 \path[draw=black, fill=black] (B2) circle[radius=0.075];
 \path[draw=black, fill=black] (B3) circle[radius=0.075];

\end{tikzpicture}}
\vspace{-0.007\textheight}
\vfill
\subcaption{}
\label{fig:interferenceHtobb}
\end{subfigure}

\caption{From left to right: vacuum diagram cut, corresponding interference diagram for $H\rightarrow gg$, and two interference diagrams for $H\rightarrow b\bar{b}$.
Red lines in figs.~\ref{fig:interferenceHtogg} and~\ref{fig:interferenceHtobb} are Cutkosky cuts that involve the observable, and dashed lines denote non-pinched threshold singularities.}
\label{fig:Htogg}
\end{figure}
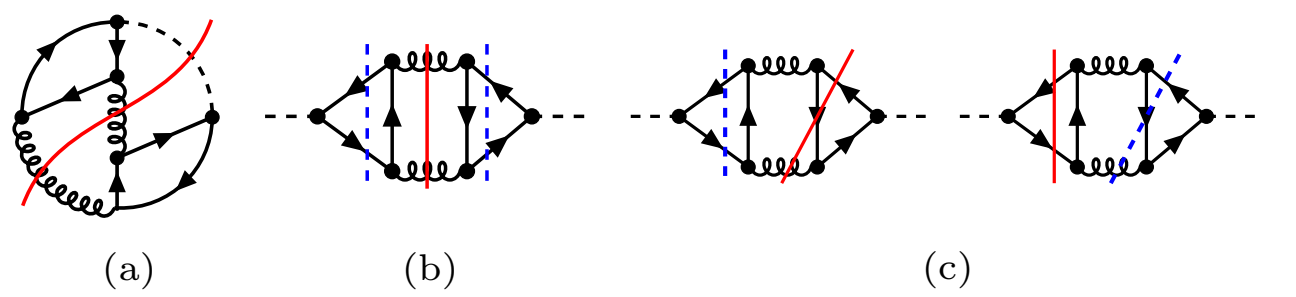

Consider as a counterexample the leading-order loop-induced decay process $H\rightarrow gg$, mediated by a closed bottom-quark loop. A single phase-space residue (fig.~\ref{fig:interferenceHtogg}) contributes to eq.~(R.4). 
Since $m_H > 2 m_b$, the diagram of fig.~\ref{fig:interferenceHtogg} has non-pinched threshold singularities associated with the production of two real bottom quarks. 
These thresholds cannot 
locally match against any other phase-space residue, since the 
cross-section only receives contributions from a single phase-space residue to begin with. 
In processes with final-state jets and kinematical cuts, the matching of \emph{non-pinched} thresholds crucially depends on both the observable and phase-space point, whereas the matching of \emph{pinched} threshold is guaranteed by the observable infrared-safety.
In particular, eq.~(R.17) does not hold in the presence of general observables for non-pinched thresholds.
Furthermore, when the matching does occur, it generally involves phase-space residues with \emph{different} multiplicities. This is the case for the two cut contributions shown in fig.~\ref{fig:interferenceHtobb}, part of the N$^2$LO QCD correction to the decay process $H\rightarrow b\bar{b}$. 

\nocite{*}
\newpage

\footnotetext{All equations marked as ``eq.~(R.x)'' in our comment refer to equations in \texttt{PRL 133.211901}~\cite{Ramirez-Uribe:2024rjg}.}

\footnotetext{
We obtain eq.~\eqref{WrongResidue} by computing the Loop-Tree Duality expression~\cite{Catani:2008xa,Bierenbaum:2010cy,PhysRevLett.122.111603,Capatti:2019ypt,Capatti:2020ytd,Capatti:2022mly}, which is a rational function $f^{\text{LTD}}\left(\{E_e\}_{e=1}^6;\{\vec{\ell}_j\cdot \vec{\ell}_i\}_{i,j=1}^3\right)$ of spatial dot products between $\vec{\ell}_1,\vec{\ell}_2$ and $\vec{\ell}_3$, and on-shell energies $E_e=\left|\vec{q}_e (\{\vec{\ell}_j\}_{j=1}^3 )\right|$. When relabelling the momenta according to fig.~\ref{fig:vacuum_diag_labelled} with $\{\ell_1,\ell_2,\ell_3\}=\{p_1,p_2,k_1\}$, $f^{\text{LTD}}$ is defined as:
$$f^\text{LTD}=\int \left[\prod_{j=1}^3 \frac{\mathrm{d}\ell_j^0}{2\pi \mathrm{i}}\right] \frac{N\left (\{\ell_j\cdot \ell_i\}_{i,j=1}^3\right )}{\prod_{e=1}^6\left (q_e^0 (\{\ell_j^0\}_{j=1}^3 )\right)^2-E_e^2+i\varepsilon},$$
where the numerator $N$ is obtained by applying Feynman rules to the vacuum graph. 

We then use eq.~(R.5) of the author's letter to compute $\mathcal{A}_D^{(\Lambda,R)}$ as:
$$
\mathcal{A}_D^{(\Lambda,R)} = \lim_{E_1\rightarrow -\lambda_{234}}\left[ 
\frac{(2 E_1)(2 E_2)}{2} \lambda_{1234}\;f^\text{LTD}
\right],
$$
with $\lambda_{i_1 \ldots i_n}=E_{i_1}+\ldots+E_{i_n}$. This yields the incorrect expression for the interference contribution that we report in eq~\eqref{WrongResidue}, omitting coupling constants, colour, and averaging factors.
}
\bibliography{apssamp}

\end{document}